\documentclass[twocolumn,pre,superscriptaddress,floatfix]{revtex4-1}
\usepackage{natbib}
\usepackage{amsmath}
\usepackage{amssymb}
\usepackage{graphicx}
\usepackage{amsbsy}
\usepackage{subfigure}
\usepackage{bm}
\usepackage{definitions}
\usepackage{localdefs}
\usepackage{rotating}

\begin{document}

\date{\today} 

\title{The head leads the body: a curvature-based kinematic description of
  \Celegans}

\author{Venkat Padmanabhan}
\affiliation{Department of Mechanical Engineering, Texas Tech University, Lubbock, Texas, United States of America}

\author{Zeina S. Khan} 
\author{Deepak E. Solomon}
\affiliation{Department of Chemical Engineering, Texas Tech University, Lubbock, Texas, United States of America}

\author{Andrew Armstrong}
\author{Kendra P. Rumbaugh}
\affiliation{Department of Surgery, Texas Tech University Health Sciences Center, Lubbock, Texas, United States of America}

\author{Siva A. Vanapalli}
\affiliation{Department of Chemical Engineering, Texas Tech University, Lubbock, Texas, United States of America}

\author{Jerzy Blawzdziewicz}
\affiliation{Department of Mechanical Engineering, Texas Tech University, Lubbock, Texas, United States of America}


\begin{abstract}
\CElegans, a free-living soil nematode, propels itself by producing
undulatory body motion and displays a rich variety of body shapes and
trajectories during its locomotion in complex environments. Here we
show that the complex shapes and trajectories of \Celegans\ have a
simple analytical description in curvature representation. Our model
is based on the assumption that the curvature wave is generated in the
head segment of the worm body and propagates backwards. We have found
that a simple harmonic function for the curvature can capture multiple
worm shapes during the undulatory movement. The worm body trajectories
can be well represented in terms of piecewise sinusoidal curvature
with abrupt changes in amplitude, wavevector, and phase.
\end{abstract}

\maketitle

\section*{Author Summary}

A thorough understanding of the kinematics of \Celegans\ is crucial to
investigate complex behaviors such as locomotion, chemotaxis and
thermotaxis. In this work, we show that \Celegans\ propels itself and
turns using a simple set of elementary movements, each of which
corresponds to a single harmonic curvature wave. The curvature wave
that originates in the head segment propagates along the worm body,
and abrupt changes in the amplitude, wavevector, and phase of the wave
result in the corresponding changes of the overall direction of a
moving nematode. We demonstrate that our model can be used not only to
describe the individual worm body postures but also to determine the
entire trail with its multiple shapes.  A deeper insight into the
repertoire of simple elementary movements of \Celegans\ will advance
qualitative studies of nematode behavior.  We also expect that it will
enable us to decipher the neural-sensory control of the nematode.

\section*{Introduction}
\label{Introduction}

\subsection*{Motivation}

A one-millimeter long soil dwelling nematode \CElegans\ is a model
organism for comprehensive investigations spanning genetics
\cite{Fire-Xu-Montgomery-Kostas-Driver-Mello:1998,
  Grishok-Tabara-Mello:2000}, neural control
\cite{White-Southgate-Thomson-Brenner:1986,
  Chen-Hall-Chklovskii:2006}, sensory transduction
\cite{Bargmann-Kaplan:1998},
behavior
\cite{Yen-Wyart-Xie-Kawai-Kodger-Chen-Wen-Samuel:2010,
  Pierce-Shimomura-Morse-Lockery:1999} and locomotion
\cite{Grey-Lissmann:1964,
  Karbowski-Cronin-Seah-Mendel-Cleary-Sternberg:2006}.  With a total
of 959 nongonadal cells, of which exactly 302 are neurons
\cite{White-Southgate-Thomson-Brenner:1986}, \Celegans\ is a very
simple organism.  Yet, it can efficiently move in complex environments
\cite{Sawin-Ranganathan-Horvitz:2000, Qin-Wheeler:2007,
  Park-Hwang-Nam-Martinez-Austin-Ryu:2008, Jung:2010}, purposefully
adjust its behavior to mechanical 
\cite{Shen-Arratia:2011,%
  Sznitman-Shen-Purohit-Arratia:2010,%
  Berri-Boyle-Tassieri-Hope-Cohen:2009}, chemical 
\cite{Faumont-Miller-Lockery:2005,%
  Pierce-Dores-Lockery:2005}, and thermal stimuli
\cite{Goodman-Schwarz:2003,
  Chalfie-Sulston-White-Southgate-Thomson-Brenner:1985,
  Matsuoka-Gomi-Shingai:2008, Ramot-MacInnis-Lee-Goodman:2008,
  Nakazato-Mochizuki:2009}, and learn 
\cite{Goodman-Schwarz:2003,%
Beale-Li-Tan-Rumbaugh:2006}.

So far many studies of nematode behavior have focused on gross
characteristics, such as the probability of turns or the fraction of
worms that approach a chemical signal within a prescribed time
\cite{Beale-Li-Tan-Rumbaugh:2006,%
Bargmann-Hartwieg-Horvitz:1993,%
Kaplan-Badri-Zachariah-Ajredini-Sandoval-Roje-Levine-Zhang-Robinette-Alborn-Zhao-Stadler-Nimalendran-Dossey-Brueschweiler-Vivanco-Edison:2009%
}.  
Also, typical body postures of
\Celegans\ (e.g., sinusoidal undulations, shallow turns, and sharp
$\Omega$- or loop-turns) have been classified and used to characterize
worm trajectories.  Such descriptions of worm motion, however, are
inadequate because they do not address more advanced questions
regarding worm behavior: for example, how the nervous, sensory, and
motor systems of \Celegans\ control its body to generate efficient
propulsion and move the nematode towards a food source or away from
danger. To gain insight into such complex behaviors a quantitative
approach to model worm dynamics is a prerequisite.

\begin{figure}[b]
\begin{picture}(270,180)
\putpic{50,0}{
\putpic{0,0}{
\includegraphics[width=\figsizeThird]{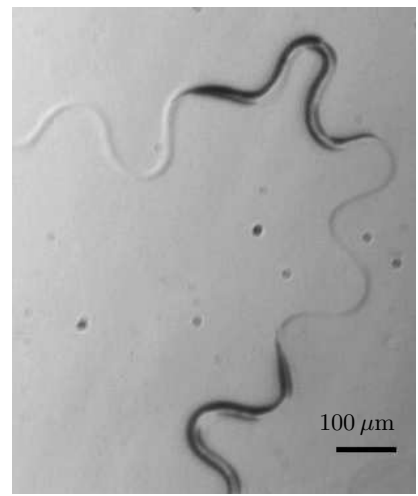}
}
\putpic{116,24}{100$\,\mu$m}
}
\end{picture}

\caption{{\bf Trail and body postures of \Celegans\ crawling on agar.}
  The figure shows two superimposed snapshots of the nematode crawling
  on a 4 \textit{wt}\% agar surface.}
\label{Track 6 snap}
\end{figure}

An important contribution to develop quantitative models of worm gait
has been made by Stephens \textit{et al.\/}
\cite{Stephens-Kerner-Bialek-Ryu:2008}.  They have shown that a set of
complex shapes that \Celegans\ assumes during crawling has a simple
representation in orientational coordinates intrinsic to the worm
body.  In their model a significant fraction of such shapes are
expressed in terms of several eigenmodes of a correlation matrix for
orientation of body segments.  This "eigenworm" description provides a
low dimensional representation of the complex space of worm postures,
which is a crucial step towards developing realistic models of the
worm behavior.

On the other hand, the eigenworm picture
\cite{Stephens-Kerner-Bialek-Ryu:2008} is not intuitive, and it misses
an important characteristic of worm trajectories, recently described
by Kim \textit{et al.}  \cite{Kim-Park-Mahadevan-Shin:2011}.  Based on
an analysis of trajectory images, they have proposed
\cite{Kim-Park-Mahadevan-Shin:2011} that shallow turns of crawling
worms can be described using a model where the worm performs simple
sinusoidal undulations with suddenly changing parameters such as
wavelength and amplitude.  While this model agrees well with
experimental observations of \Celegans\ performing low-amplitude
undulations \cite{Kim-Park-Mahadevan-Shin:2011}, it is not sufficient
for a description of $\Omega$-turns and other sharp turns.

\input{figtex/schematic}

\subsection*{Piecewise-harmonic-curvature  model}

Our curvature-based description of \Celegans\ motion is motivated by
geometrical observations of trails of nematodes moving without slip on
a soft agar substrate.  In this work we build on the ideas of
\cite{Stephens-Kerner-Bialek-Ryu:2008} and
\cite{Kim-Park-Mahadevan-Shin:2011} to develop a simple
curvature-based description of worm kinematics, obtaining a model that
overcomes the shortcomings described above.  As in
\cite{Kim-Park-Mahadevan-Shin:2011}, we analyze entire trails of
crawling nematodes (rather than their individual body shapes) to
capture sudden changes of the gait that the worm uses to navigate its
environment.  However, we formulate our description in terms of
intrinsic geometric quantities (arc length $\arcLength$ and local
curvature $\curvature$), rather than real-space coordinates, and this
allows us to go beyond describing only shallow turns and low-amplitude
undulations.  As a result, we obtain a low-dimensional representation
of all worm postures and trajectories in terms of uncomplicated basis
functions.

A typical trail is shown in Fig.\ \ref{Track 6 snap}, where two images
are superimposed to depict the worm postures at different times.  The
worm makes deep U-shaped undulations throughout the trajectory.  The
depth of the undulations varies in time, and due to these variations
the overall direction of the worm motion changes.  The worm
occasionally assumes an $\Omega$-like shape (as seen in the top right
corner of Fig.\ \ref{Track 6 snap}), which results in a significant
change of direction.

\paragraphP{Assumptions}

Our piecewise-harmonic-curvature (PHC) model is built on the following
three assumptions: (i) the worm controls its body posture by
propagating the curvature backwards along the body, (ii) the curvature
wave is generated in the front (head) segment of the worm body; the
time variation of the curvature of this segment serves as the boundary
condition for the curvature wave and determines this wave completely,
and (iii) the variation of the head segment curvature can be described
in terms of a simple piecewise-harmonic function.

As shown below, our model is confirmed by a detailed analysis of
trails of worms crawling without slip on agar substrate.
Trail segments described by harmonic variation of the curvature
are clearly identified, and the jumps of the curvature-wave amplitude,
wavevector, and phase from segment to segment are determined. We thus
obtain an analytic representation of the entire trail in terms of a
relatively small number of parameters.

\begin{figure}

\begin{picture}(350,350)
\putpic{25,0}{
\putpic{43,300}{
\includegraphics[scale=0.32, angle = 90]{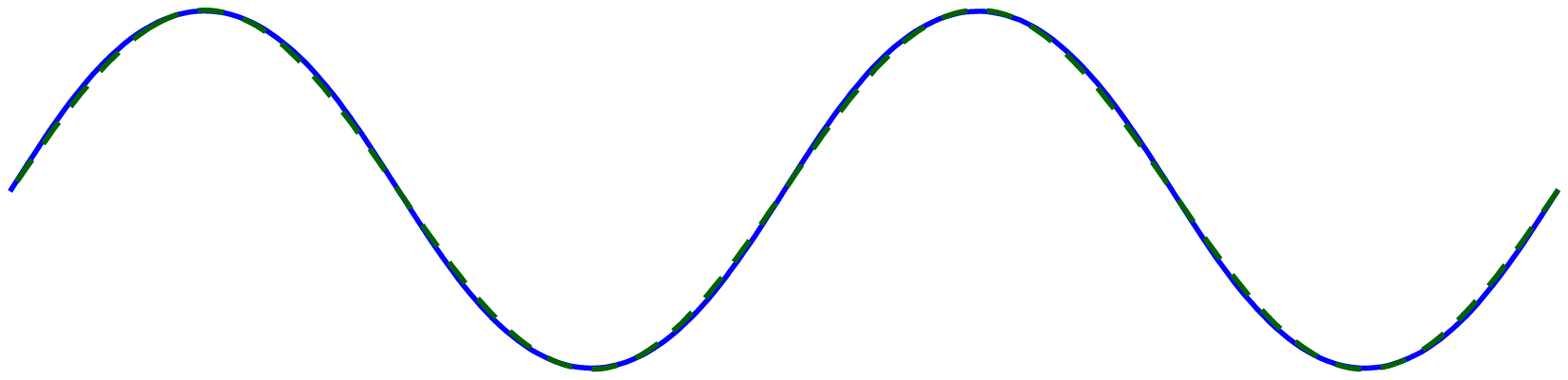}
}
\putpic{43,237}{
\includegraphics[scale=0.32, angle = 90]{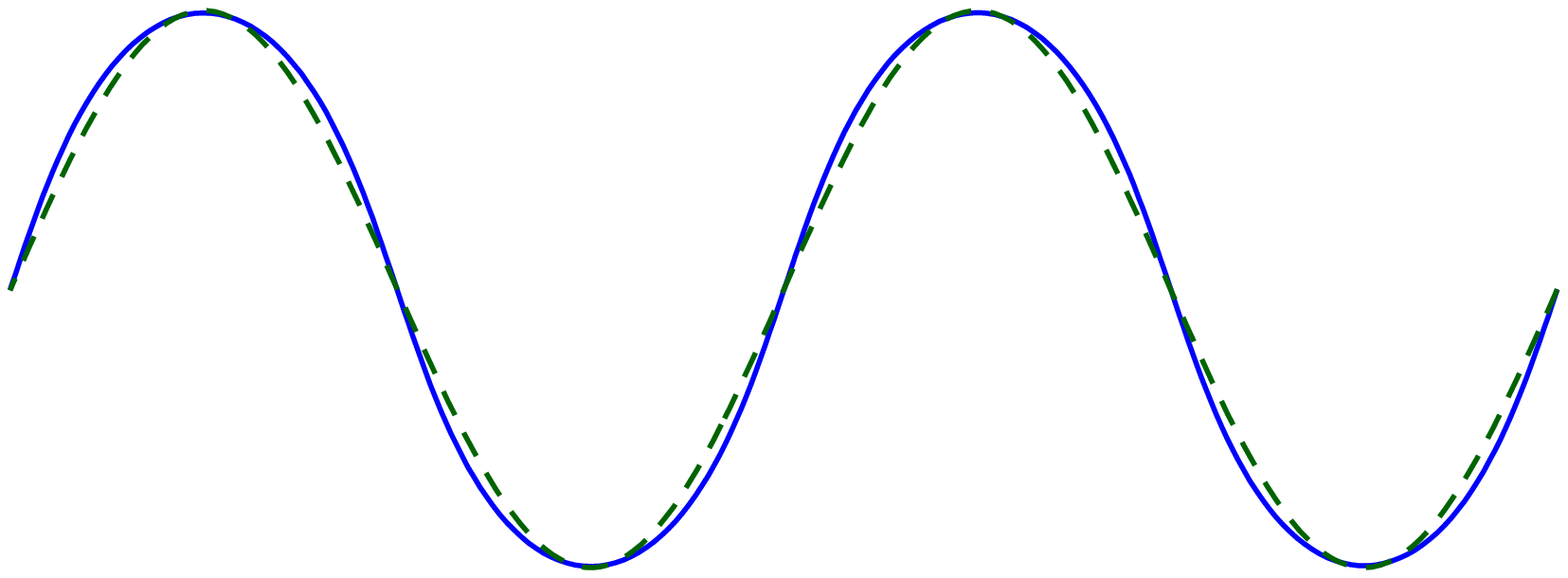}
}
\putpic{43,170}{
\includegraphics[scale=0.32, angle = 90]{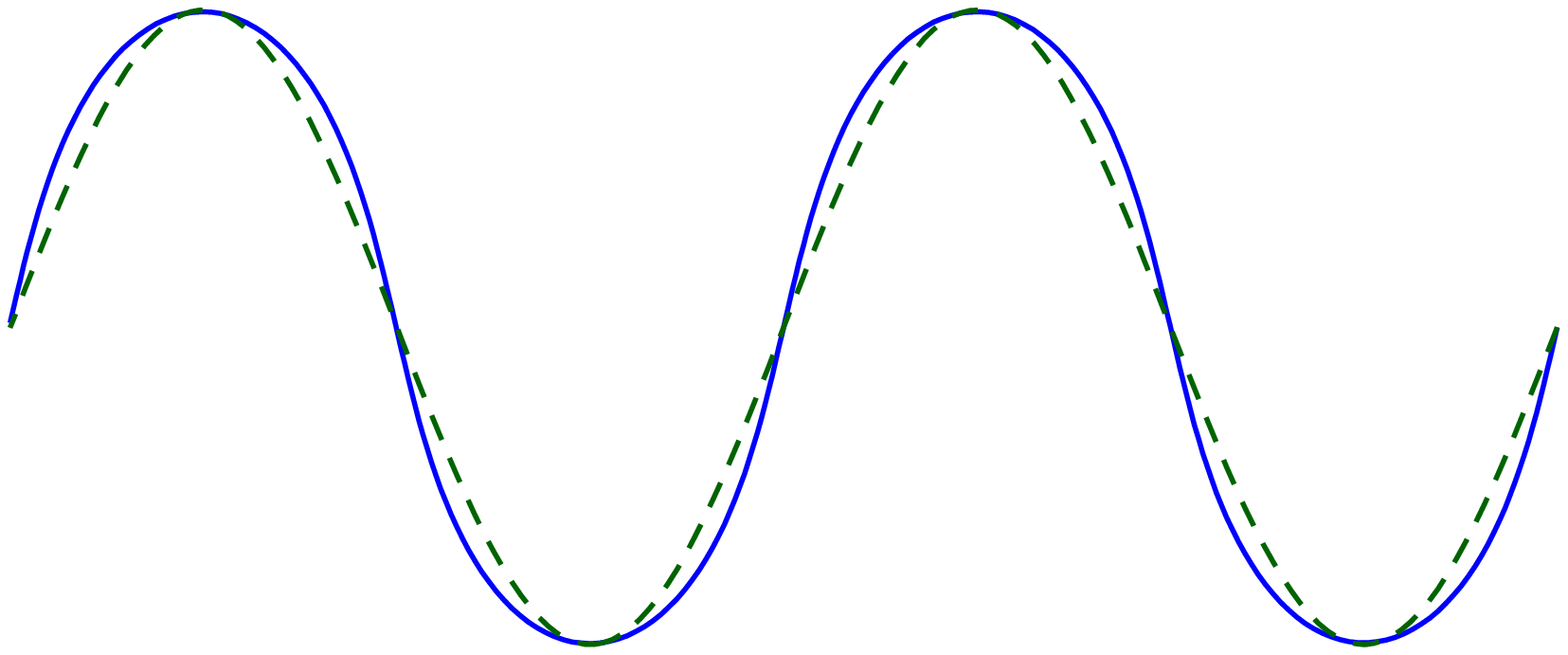}
}
\putpic{43,85}{
\includegraphics[height = 2.1in, width = 1.2in,angle = 90]{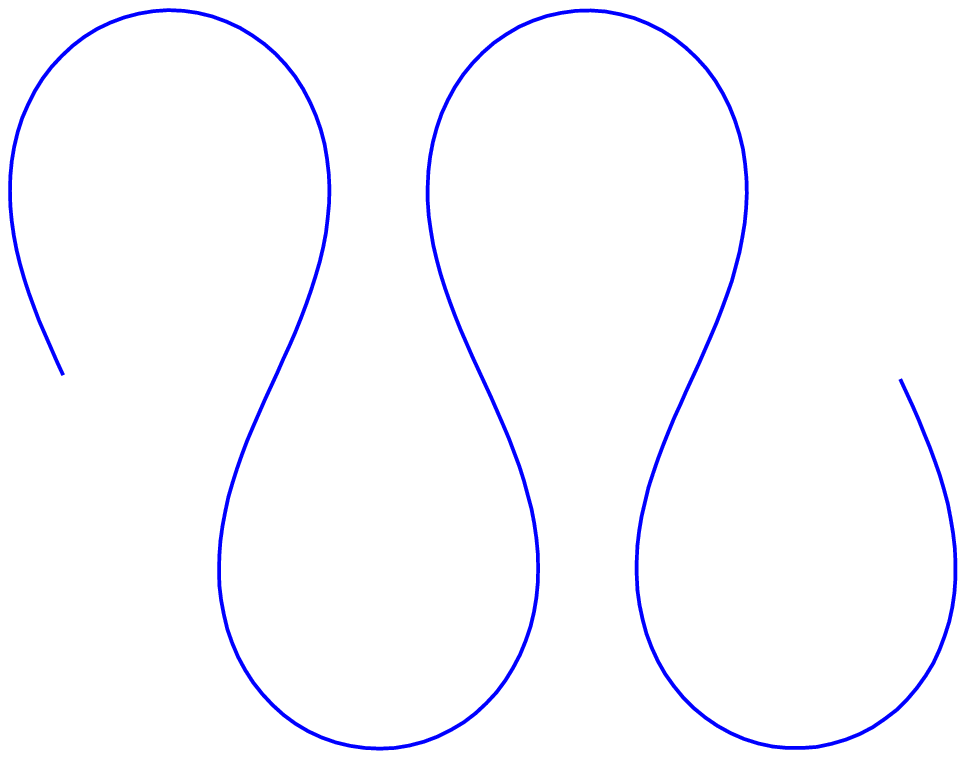}
}
\putpic{43,0}{
\includegraphics[width = 1.11in, angle = 90]{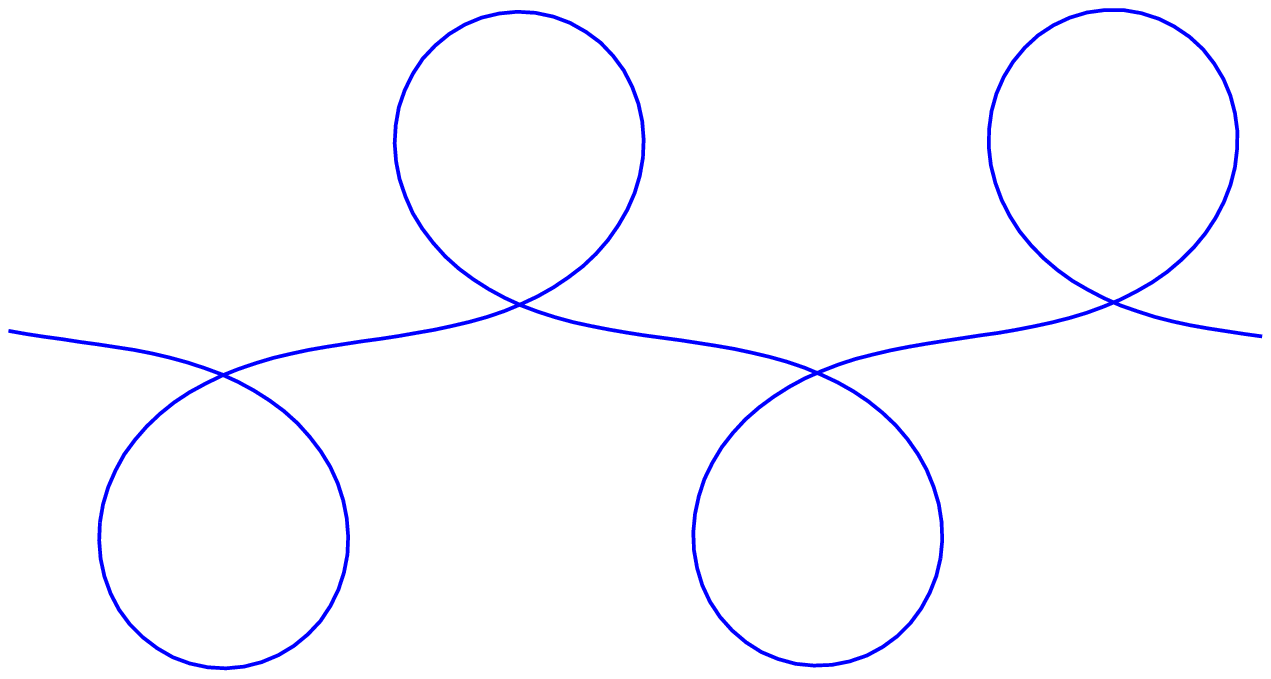}
}
\putpic{-5,315}{\small$\amplitude/$\wavevector$=0.6$}
\putpic{-5,262}{\small$\amplitude/$\wavevector$=1.0$}
\putpic{-5,199}{\small$\amplitude/$\wavevector$=1.2$}
\putpic{-5,121}{\small$\amplitude/$\wavevector$=2.0$}
\putpic{-5,39}{\small$\amplitude/$\wavevector$=3.0$}
}
\end{picture}

\caption{{\bf Family of shapes obtained from a harmonic curvature
    function.} Solid lines show the solutions of Frenet--Serret
  equations \eqref{explicit Frenet-Serret equations} with the harmonic
  curvature \eqref{harmonic curvature} with different values of
  normalized amplitude $\amplitude$/$\wavevector$.  The dashed lines
  correspond to the real-space sinusoidal form \eqref{sine
    equation}. }

\label{Sine compare}
\end{figure}

\input{figtex/worm_shapes_1}

\begin{figure*}

\begin{picture}(350,112)
\putpic{0,0}{

\putpic{10,0}{
\putpic{0,0}{
\includegraphics[width=0.22\textwidth]{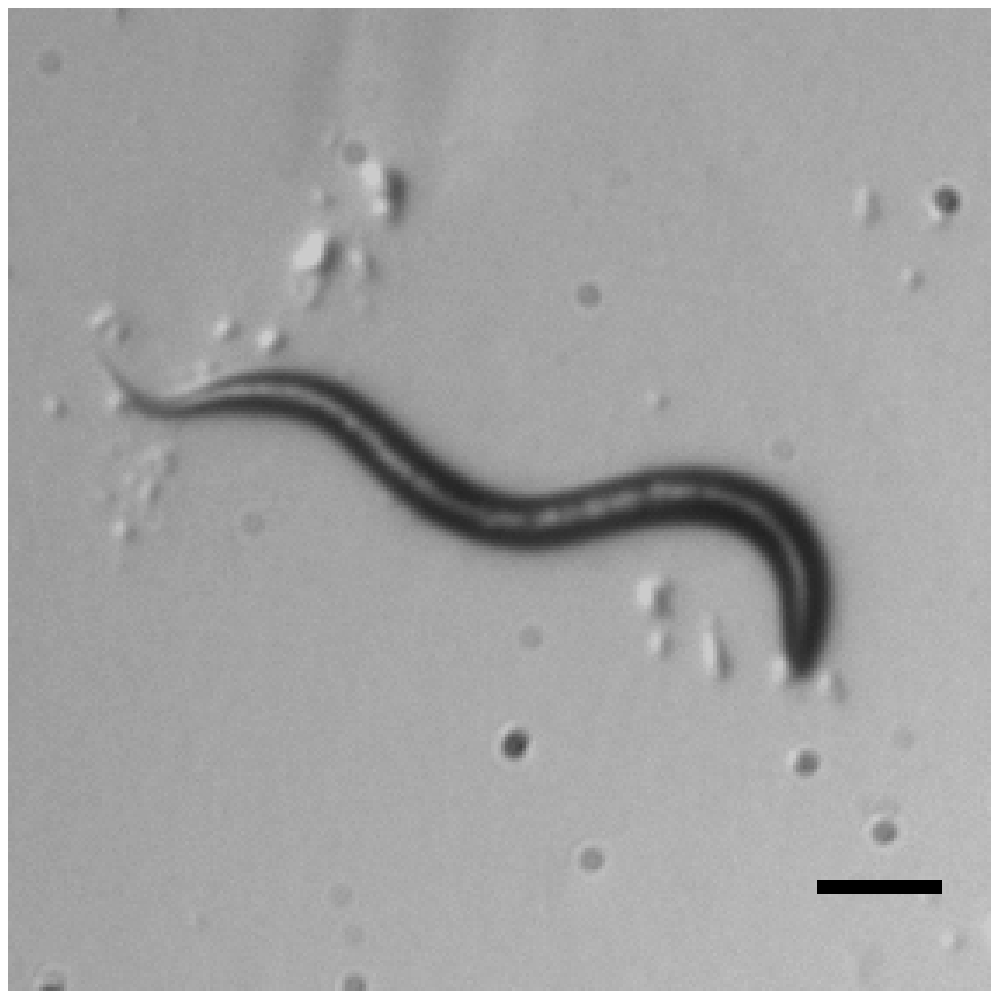}
}
\putpic{80,18}{100$\,\mu$m}
}

\putpic{125,0}{
\reflectbox{\includegraphics[width=0.22\textwidth, angle=90]{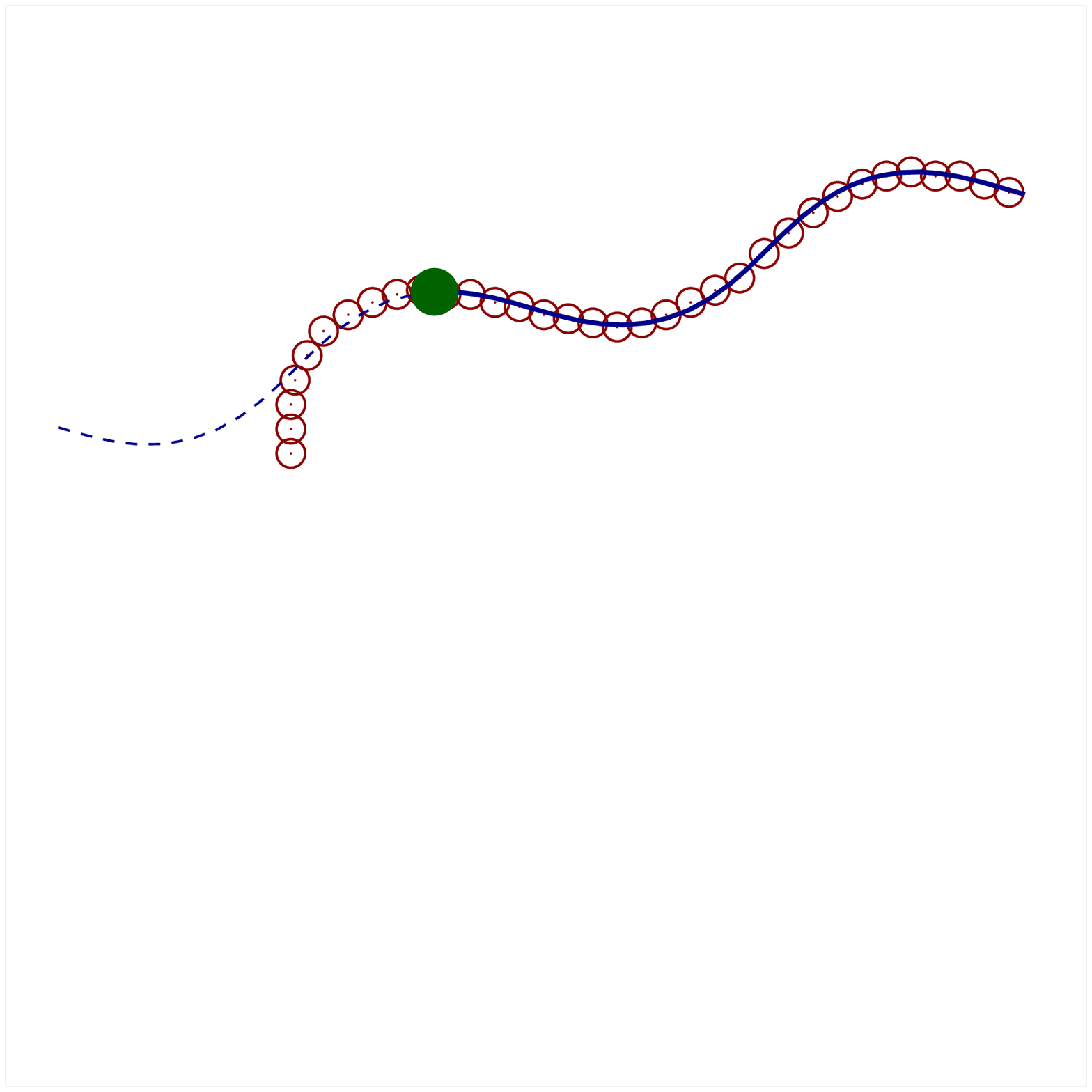}}
}
\putpic{124,65}{
\includegraphics[width=0.115\textwidth, angle=270]{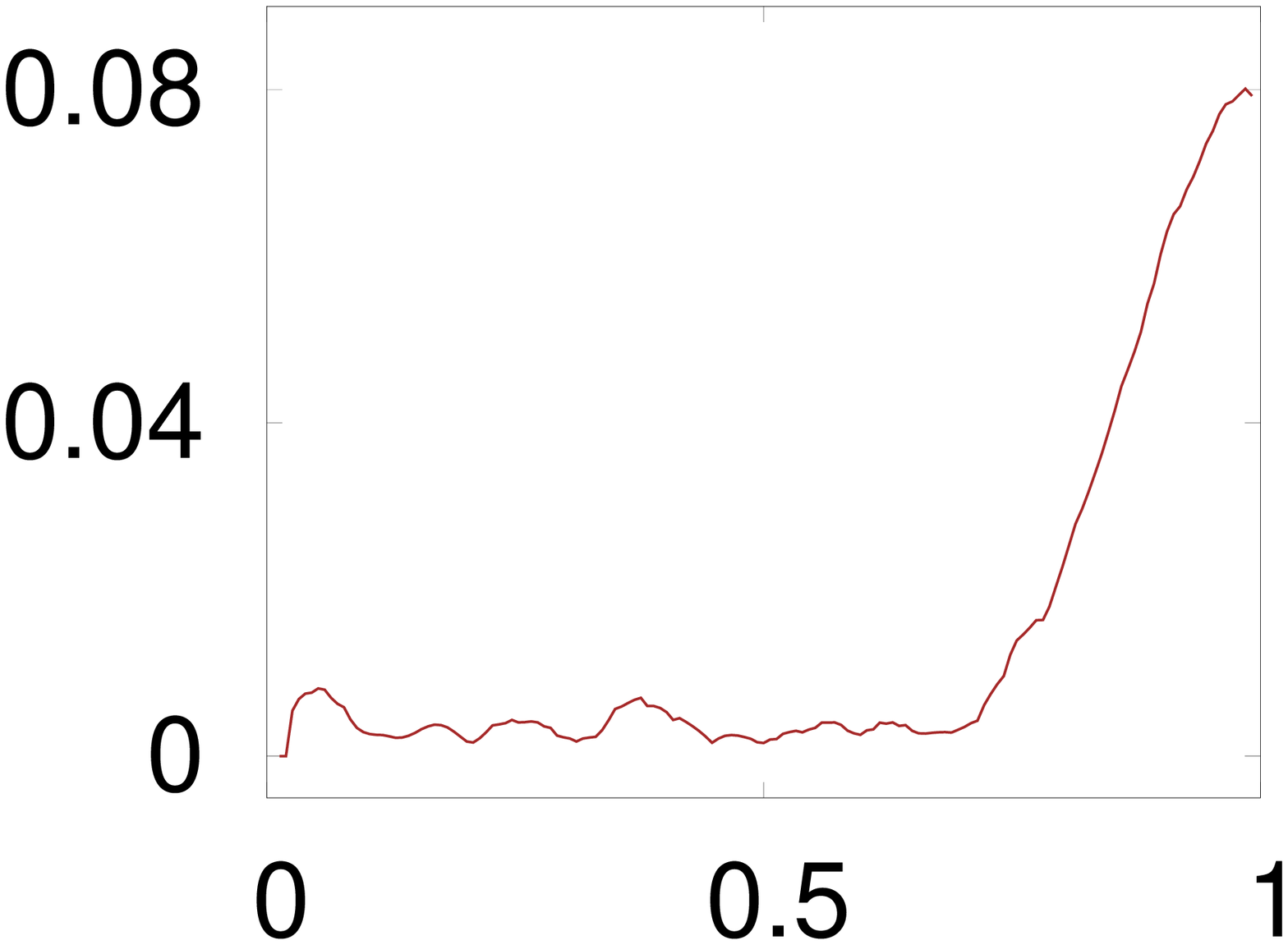}
}
\putpic{170,4}{$\arcLength/\wormLength$}
\putpic{128,40}{$\epsilon$}

\putpic{240,0}{
\reflectbox{\includegraphics[width=0.22\textwidth, angle=90]{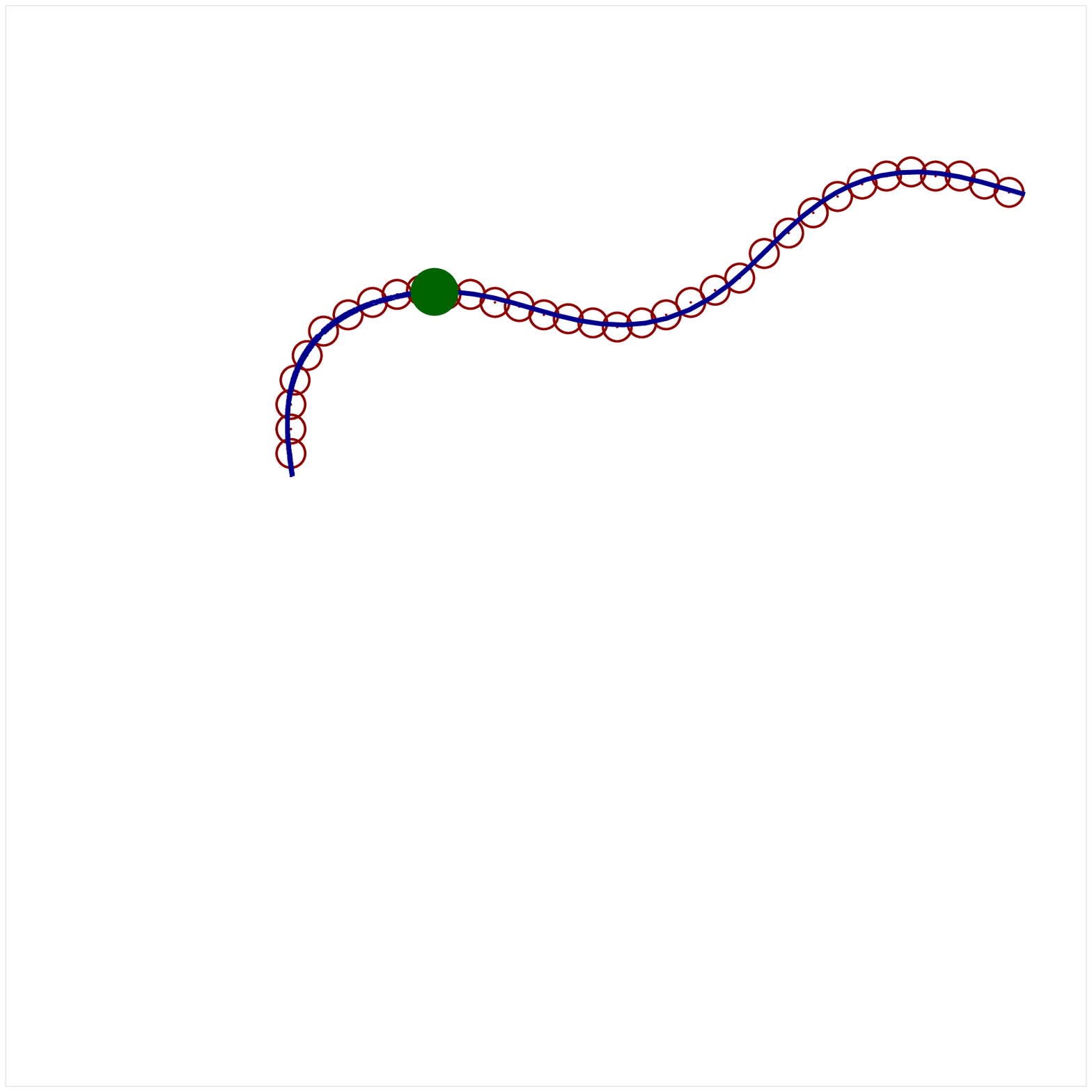}}
}
\putpic{239,65}{
\includegraphics[width=0.115\textwidth, angle=270]{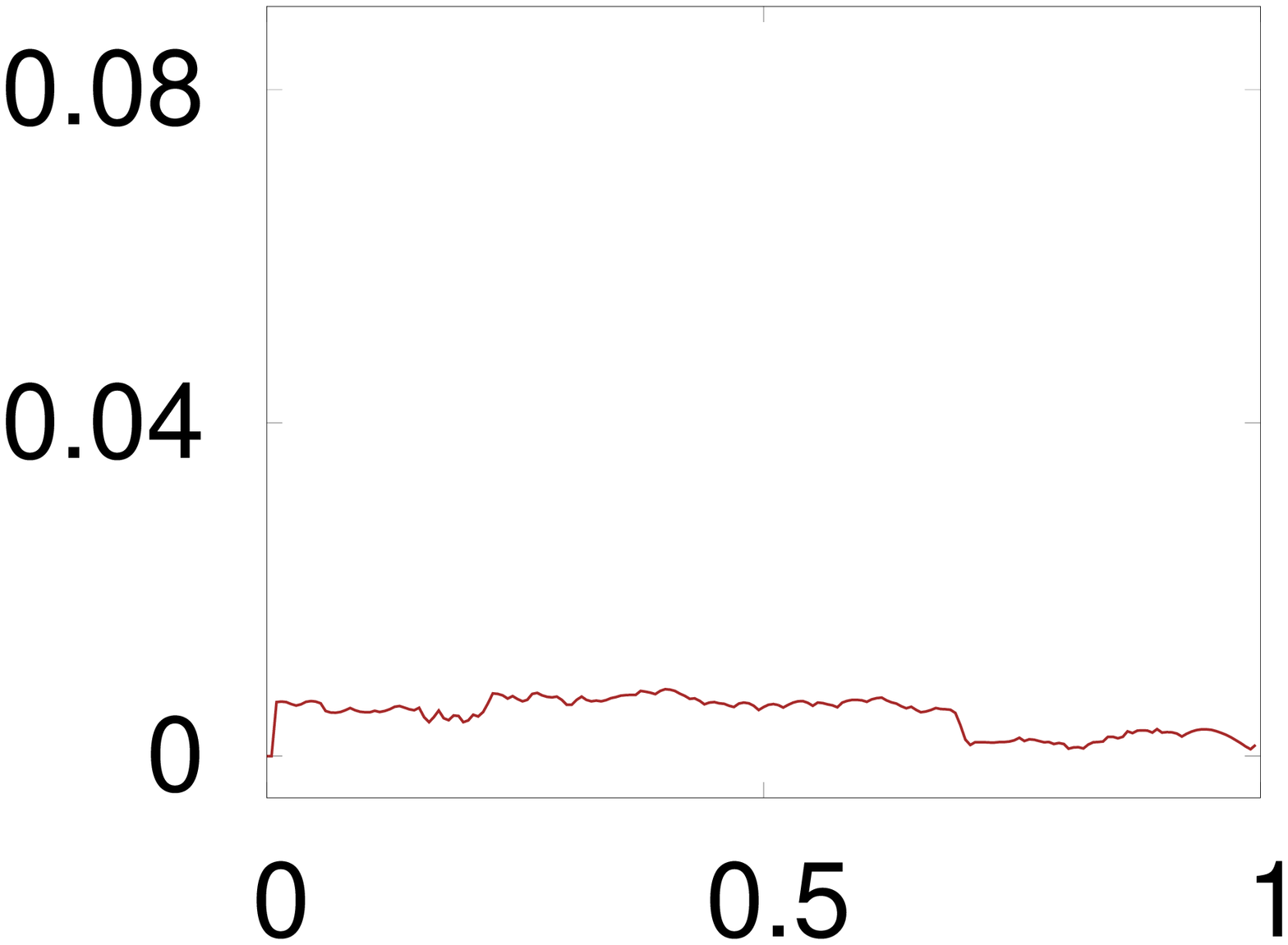}
}
\putpic{285,4}{$\arcLength/\wormLength$}
\putpic{243,40}{$\epsilon$}

\putpic{15,100}{A}
\putpic{130,100}{B}
\putpic{245,100}{C}

}
\end{picture}

\caption{{\bf Body posture of \Celegans\ with two distinct curvature
    modes.} (A) The experimental image of the worm. (B) The skeleton
  data (open circles) and the harmonic-curvature fit for the tail
  segment (solid line). The extension of the fit that does not follow
  the head segment is shown by dashed line, and the point where the
  curvature mode changes is indicated by a filled circle. (C) The best
  fit of the two-mode harmonic-curvature representation
  \eqref{harmonic curvature} (solid line).  Since the line is obtained
  by integrating second-order differential equations \eqref{explicit
    Frenet-Serret equations}, it is continuous and has a continuous
  slope.  The insets in (B) and (C) show the local fit error
  \eqref{local error function} along the skeleton of the nematode.
  For the single-mode fit (B) the error rapidly increases after the
  point indicated by the filled circle, whereas for the continuous
  two-mode fit (C), the local error is below 1\% along the whole
  body.}
\label{Worm shapes 2}
\end{figure*}

\begin{figure}

\begin{picture}(270,133)
\putpic{0,15}{

\putpic{10,0}{
\putpic{0,0}{
\includegraphics[width=0.22\textwidth]{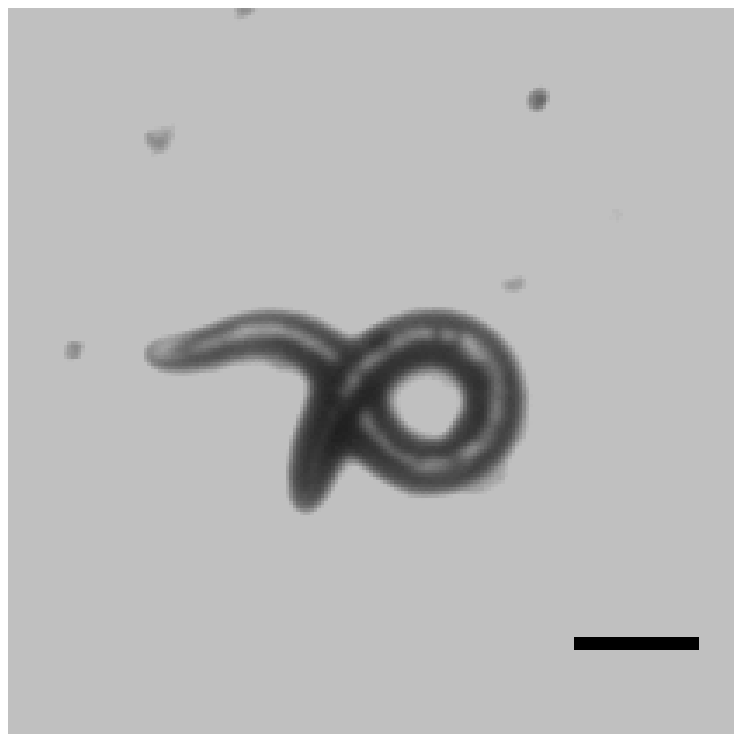}
}
\putpic{80,21}{100$\,\mu$m}
}
\putpic{121,0}{
\reflectbox{\includegraphics[width=0.22\textwidth,angle=90]{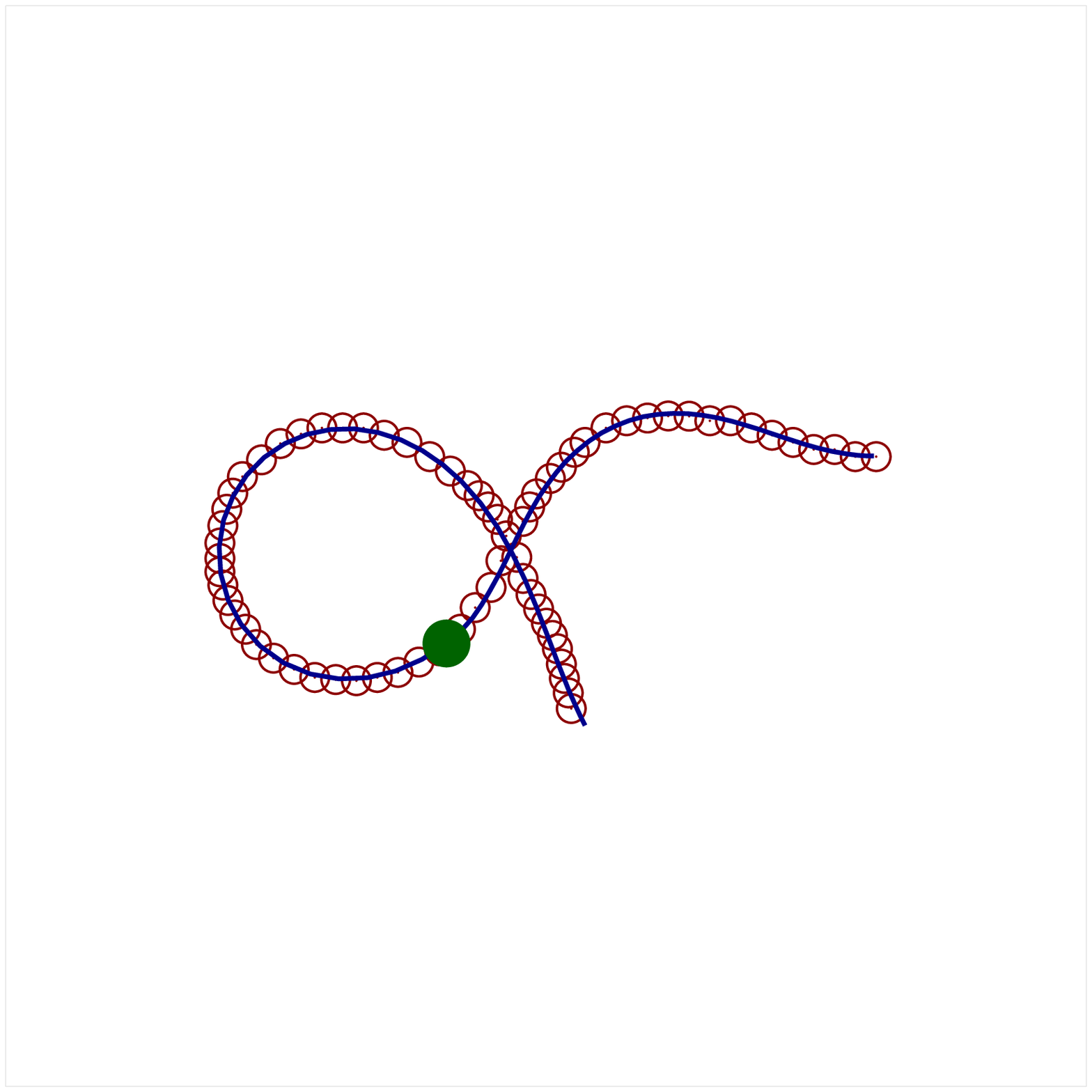}
}}

\putpic{15,100}{A}
\putpic{130,100}{B}

}
\end{picture}

\caption{{\bf Loop turn with two distinct curvature modes.} (A)
  Experimental image.  (B) The two-mode harmonic-curvature
  representation (solid line); the skeleton data (open circles); the
  location of the mode change (filled circle).}
\label{Loop turn}
\end{figure}

\input{figtex/piece_fit-trail}
\begin{figure}

\newcommand{\trackNo}[1]{{\footnotesize \uppercase{#1}}}
\newcommand{\trackNos}[1]{{\scriptsize \uppercase{#1}}}
\renewcommand{\trackNos}[1]{{\footnotesize\uppercase{#1}}}

\begin{picture}(575,290)
\putpic{0,325}{

\putpic{0,142}{

\putpic{0,0}{ 

\putpic{25,-179}{
\includegraphics[width=0.26\textwidth, angle=270]{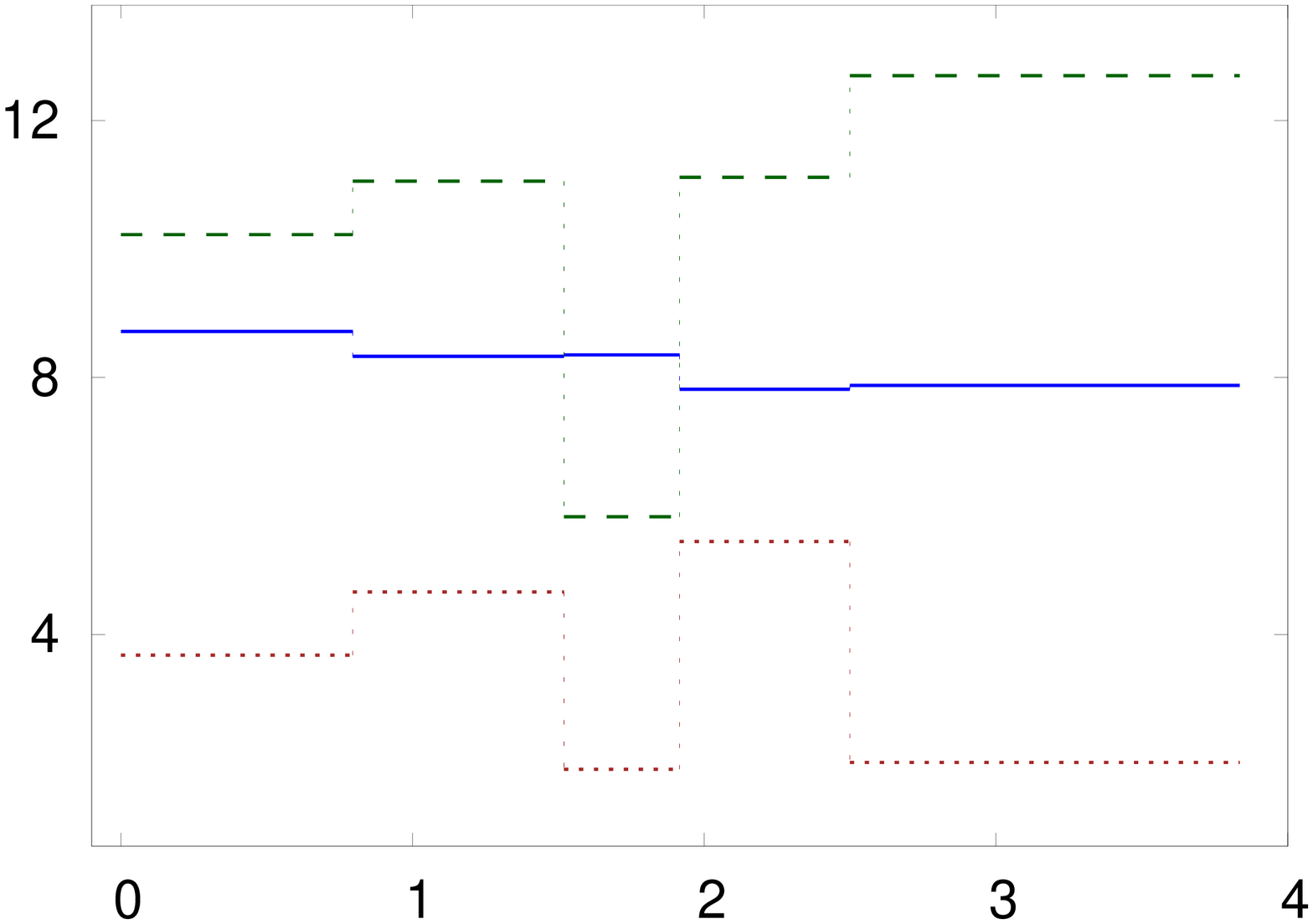}
}
\putpic{129,-310}{$\arcLength/\wormLength$}
\putpic{25,-255}{\begin{sideways}{$\wormLength\amplitude,\wormLength\wavevector,\phase$}\end{sideways}}
}

\putpic{12,-187}{A}

} 

\putpic{-213,-8}{

\putpic{0,0}{

\putpic{240,-177}{
\includegraphics[width=0.26\textwidth, angle=270]{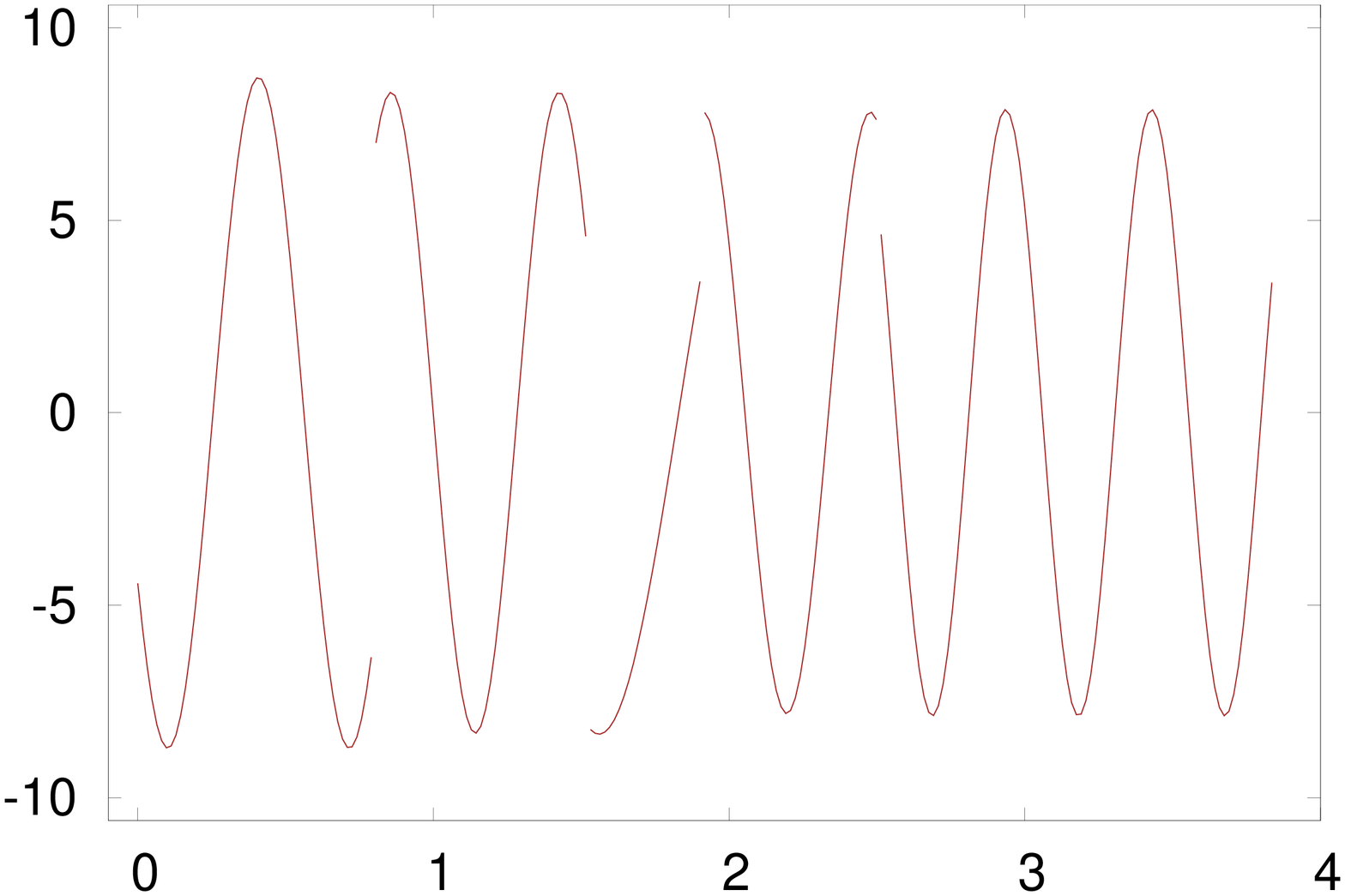}
}
\putpic{342,-309}{$\arcLength/\wormLength$}
\putpic{240,-237}{$\curvature$}

}

\putpic{225,-187}{B}

} 

}

\end{picture}
\caption{{\bf Parameters of the PHC representation.} (A) The
  normalized amplitude $\amplitude$ (blue -- solid), wavevector
  $\wavevector$ (green -- dashed), and phase $\phase$ (red -- dotted)
  of the curvature wave along the trail shown in Fig.\ \ref{Piece fit
    - trail}. (B) The evolution of curvature along the same trail.}
\label{Piece fit - parameters}
\end{figure}

\input{figtex/track_651}

\input{figtex/track_34}

\section*{Results}

\subsection*{Curvature representation of the worm trail}

It is convenient to describe  nematode trails and instantaneous
body shapes using the parametric description
\begin{equation}
\label{parametric body shape}
x=x(\arcLength), \qquad y=y(\arcLength)
\end{equation}
for the Cartesian $x$ and $y$ coordinates of all points along the trail.  The
parameter $s$ in Eq.\ \eqref{parametric body shape} represents the
arc length along the trail.   The worm body posture at time $\timet$
corresponds to a portion of the trail between the points
\begin{equation}
\label{head tail position}
\arcLengthTail=\arcLengthTail(\timet), \qquad 
\arcLengthHead=\arcLengthTail(\timet)+\wormLength,
\end{equation}
where $\arcLengthTail$ is the tail position, $\arcLengthHead$ is the
head position, and $\wormLength$ is the length of the worm.  If the
worm is moving with a constant velocity $\wormVelocity$, the tail
position is given by
\begin{equation}
\label{tail position time}
\arcLengthTail(\timet)=\arcLengthTail(0)+\wormVelocity\timet.
\end{equation}
The above description of the trail geometry is illustrated in
Fig.\ \ref{Schematic}.

The shape of the worm trail in the parametric form \eqref{parametric
  body shape} can be evaluated from the local curvature
$\kappa=\curvature(\arcLength)$ using the Frenet--Serret equations
\begin{equation}
\label{Frenet--Serret}
\frac{\diff\tangentVector(\arcLength)}{\diff\arcLength}=\curvature(\arcLength)\normalVector(\arcLength),
\end{equation}
where 
\begin{equation}
\label{tangent and normal vector}
\tangentVector=\frac{\diff x}{\diff\arcLength}\ex+\frac{\diff y}{\diff\arcLength}\ey, \qquad
\normalVector=\frac{\diff x}{\diff\arcLength}\ey-\frac{\diff y}{\diff\arcLength}\ex
\end{equation}
are the unit vectors tangent and normal to the trail, respectively.
Combining relations \eqref{Frenet--Serret} and \eqref{tangent and
normal vector} yields a set of coupled second-order differential
equations
\begin{subequations}
\label{explicit Frenet-Serret equations}
\begin{eqnarray}
\label{explicit Frenet-Serret equations x}
\frac{\diff^2 x}{\diff\arcLength^2}&=&-\curvature \frac{\diff y}{\diff\arcLength}\\
\label{explicit Frenet-Serret equations y}
\frac{\diff^2 y}{\diff\arcLength^2}&=&\phantom{-}\curvature \frac{\diff x}{\diff\arcLength}
\end{eqnarray}
\end{subequations}
for the trail shape \eqref{parametric body shape}.  Equations
\eqref{explicit Frenet-Serret equations} are solved with the initial
conditions
\begin{subequations}
\label{initial conditions for the worm body}
\begin{eqnarray}
\label{initial conditions for the worm body x}
x(\arcLength = 0)=x_0,&\qquad&\left.\frac{\diff x(\arcLength)}{\diff\arcLength}\right\vert_{\arcLength=0}=\tangentVectorComponent_{x_0},\\
\label{initial conditions for the worm body y}
y(\arcLength = 0)=y_0,&\qquad&\left.\frac{\diff y(\arcLength)}{\diff\arcLength}\right\vert_{\arcLength=0}=\tangentVectorComponent_{y_0},
\end{eqnarray}
\end{subequations}
where $(x_0,y_0)$ is the position of the worm tail at the beginning of
  the trail and the tangent unit vector,
  $\tangentVector_0=(\tangentVectorComponent_{x_0},\tangentVectorComponent_{y_0})$
  describes the orientation of the tail at $t = 0$.

\subsection*{Harmonic-curvature variation}

We argue that worm trails (and the corresponding sequences of worm
shapes) can be accurately described in terms of a simple
harmonic-curvature variation
\begin{equation}
\label{harmonic curvature}
\curvature(\arcLength)=\amplitude\sin(\wavevector\arcLength+\phase),
\end{equation}
where $\amplitude$ is the amplitude, $\wavevector$ is the wavevector,
and $\phase$ is the phase of the harmonic wave.  Below we demonstrate
that a combination of segments of the lines obtained by integration of
the curvature \eqref{harmonic curvature} with different values of the
parameters $\amplitude$, $\wavevector$, and $\phase$ can be used to
construct the entire worm trail.

The family of shapes obtained by changing the amplitude $\amplitude$
in Eq.\ \eqref{harmonic curvature}, with $\wavevector$ and $\phase$
fixed, is depicted in Fig.\ \ref{Sine compare}. Other shapes,
corresponding to different values of $\wavevector$ and $\phase$, can
be obtained from this family by affine transformations.  A comparison
of the lines plotted in Fig.\ \ref{Sine compare} with the worm tracks
and body shapes shown in Figs.\ \ref{Track 6 snap} and \ref{Worm
shapes 1}--\ref{Piece fit 651} reveals close similarity.  For example,
the shape of the line with the dimensionless amplitude
$\amplitude/\wavevector=2$ closely resembles the $\Omega$-turns made
by the worm in Figs.\ \ref{Worm shapes 1}C and
\ref{Piece fit 651}.  Similarly, for $\amplitude/\wavevector=3$, the
shape resembles the loop turn as seen in Fig.\ \ref{Loop turn}.

In Fig.\ \ref{Sine compare} the curvature-based
harmonic-wave description \eqref{harmonic curvature} is compared with
the real-space harmonic wave
\begin{equation}
\label{sine equation}
y=\amplitudeY\sin(\wavevectorX x+\phaseX)
\end{equation}
that was used in \cite{Kim-Park-Mahadevan-Shin:2011} to describe worm
trails with shallow turns. For low amplitudes of the curvature wave
$\amplitude/\wavevector\lesssim 1$ the two relations \eqref{harmonic
curvature} and \eqref{sine equation} yield nearly equivalent shapes,
but the results significantly differ outside this range.  An analysis
of the skeletonized worm shapes depicted in Fig.\ \ref{Worm shapes 1}
shows that our harmonic-curvature model
\eqref{harmonic curvature} describes nematode body postures accurately for all
amplitudes. The real-space harmonic description \eqref{sine equation}
  becomes inaccurate already for moderate-amplitude undulation
  (cf. Fig.\ \ref{Worm shapes 1}B).

\subsection*{Piecewise-harmonic-curvature function}
\label{Piece-wise description}

Our simple harmonic-curvature model \eqref{harmonic curvature} can be
applied to describe both individual worm postures and entire nematode
trails (such as the trail depicted in Fig.\ \ref{Track 6 snap}).
Complete trails that include gradual and deep turns can be represented
with high accuracy using a piecewise-harmonic-curvature function.  We
first discuss piecewise-harmonic representation for individual
nematode postures, and then we analyze trajectories of crawling worms.

\paragraphP{Worm body postures}
\label{Worm shape model}

The shape of \Celegans\ shown in Fig.\ \ref{Worm shapes 2} has a
distinctly higher curvature in the head section than in the tail
section.  For such a shape, a single-mode sinusoidal-curvature
representation is insufficient.  Figure \ref{Worm shapes 2}B shows a
line obtained by integrating Frenet--Serret equations \eqref{explicit
Frenet-Serret equations} with the harmonic curvature \eqref{harmonic
curvature} and parameters adjusted to fit the tail part of the
worm. From the tip of the tail up to the point indicated by the filled
circle, the line matches the skeletonized nematode posture very
well. Beyond this point a sudden deviation occurs, as determined by
the local error shown in the inset.  We conclude that the worm
exhibits more than one mode along its body length, and there is a
sudden change in the curvature of the worm's body at the point marked,
where the transition occurs.

To accurately describe the entire worm posture, we use the
piecewise-harmonic curvature
\begin{equation}
\label{harmonic curvature 2}
\curvature(\arcLength)=\left\{\begin{matrix}\amplitudea\sin(\wavevectora\arcLength+\phasea); \qquad \arcLengthTail \le \arcLength \le \arcLengthSega \\
\amplitudeb\sin(\wavevectorb\arcLength+\phaseb); \qquad \arcLengthSega \le \arcLength \le \arcLengthHead. \\
\end{matrix}\right.
\end{equation}
Here $\arcLengthSega$ is the intermediate point at which a sudden
transition between harmonic modes occurs.  As demonstrated in
Fig.\ \ref{Worm shapes 2}C, the two-mode curvature
function yields an accurate description of the shape of the whole worm
from tail to head.

Two-mode fits of similar quality have been obtained for other nematode
shapes, such as the loop posture depicted in Fig.\ \ref{Loop turn}.
At the switching point $\arcLengthSega$ the worm curvature is often
discontinuous, but the worm contour obtained via integration of the
second-order Frenet--Serret differential equation \eqref{explicit
Frenet-Serret equations} is continuous and has a continuous
orientation.

\paragraphP{Worm trails}
\label{Worm trajectory model}

For \Celegans\ crawling without slip on a solid surface, the
instantaneous body postures can be treated as segments of the entire
worm trail (cf. Fig.\ \ref{Track 6 snap}).  Therefore, we assume that
the abrupt changes in the parameters of the curvature wave that
propagates throughout the worm body are reflected in the overall trail
shape.

A typical example of a trail made by \Celegans\ crawling on an agar
surface is depicted in Fig.\ \ref{Piece fit - trail}A.  The picture
was obtained by superimposing consecutive frames of a video recording
of a crawling worm, as described in the Methods section.  The worm
undergoes a small amount of slip, which is reflected in occasional
trail imperfections. These imperfections are also due to the random
movement that the head of the nematode makes before deciding on the
path of its motion. However, once the head continues to move in a
certain direction, the rest of the body follows the head on a
determined path.

The trail of a gradually turning worm shown in Fig.\ \ref{Piece fit -
  trail}A cannot be represented using a single harmonic-curvature
  mode.  However, similar to the results for worm postures (cf.\
  Fig.\ \ref{Worm shapes 2}), individual segments of the trail are
  consistent with the harmonic-curvature representation, as
  illustrated in Fig.\ \ref{Piece fit - trail}B.  In a certain range
  of $\arcLength$ each local harmonic-curvature fit follows the track
  with high accuracy, and then rapidly deviates from the track as
  shown in the insets in Fig.\ \ref{Piece fit - trail}B. The errors of
  fits are less than 1\% up to the mode transition point and then show
  a steep increase.
  
Based on the above observations, we expect that the entire trail can
be well represented by a continuous line corresponding to the
piecewise-harmonic curvature
\begin{equation}
\label{harmonic curvature multiple}
\curvature(\arcLength)=\left\{\begin{matrix}
   \amplitudea\sin(\wavevectora\arcLength+\phasea); 
         \qquad \arcLengthBeg \le \arcLength \le \arcLengthSega \\
   \amplitudeb\sin(\wavevectorb\arcLength+\phaseb); 
         \qquad \arcLengthSega \le \arcLength \le \arcLengthSegb \\
\,\,\vdots
   \hphantom{\sin(\wavevectorb\arcLength+\phaseb);\
        qquad \arcLengthSega \le \arcLength \le \arcLengthSegb} \\
\amplitudei\sin(\wavevectori\arcLength+\phasei); \qquad \arcLengthSegi \le \arcLength \le \arcLengthEnd. \\
\end{matrix}\right.
\end{equation}
Here $\arcLengthSega$, $\arcLengthSegb$, ... $\arcLengthSegi$ denote
the intermediate points of the worm's trail where a sudden transition
between modes occurs, and $\arcLengthBeg$ and $\arcLengthEnd$ are the
beginning and end points of the trail, respectively.  The continuous
PHC representation for the whole trail is depicted in Fig.\ \ref{Piece
fit - trail}C. The inset in Fig.\ \ref{Piece fit - trail}C shows the
local deviation of our PHC model from the skeleton data.  The low
deviation (less than 1\% with occasional jumps to about 2\%) for all
values of the arc-length parameter $\arcLength$ indicates that our
model fits the data with high accuracy.

Figure \ref{Piece fit - parameters}A shows the discontinuous variation
in the curvature-wave amplitude, wavevector and phase along the worm
trail depicted in Fig.\ \ref{Piece fit - trail}A.  Jumps of these
three parameters occur at irregular intervals, and the size of the
jumps is also irregular.  The corresponding evolution of the curvature
is illustrated in Fig.\ \ref{Piece fit - parameters}B. The curvature
is discontinuous at each mode shift, but the final real-space PHC
representation, obtained via integration of the second order
differential equation \eqref{explicit Frenet-Serret equations}, is
continuous.

Further examples of worm trails and the corresponding PHC
representations are shown in Figs.\ \ref{Piece fit 651} and \ref{Piece
  fit 34}.  Our description captures well both the gradual changes of
the overall path direction (cf.\ Figs.\ \ref{Piece fit - trail} and
\ref{Piece fit 34}) and rapid changes such as $\Omega$-turns and other
deep turns (cf.\ Fig.\ \ref{Piece fit 651}).  An analysis of the
variation in the curvature-wave parameters indicates that a deep turn
may result from an increase of the wave amplitude, a decrease of the
wave vector, or from both such changes occurring at the same time.

The nematode \Celegans\ changes suddenly its undulation mode not only
during forward motion, but also in the backward crawling gait.  It has
been observed \cite{Zhao-Khare-Feldman-Dent:2003} that upon reversal
of its direction of motion \Celegans\ often retraces part of its
track, and then abruptly veers from the previous path.  This deviation
has a very similar geometry to the abrupt divergence between the trail
of a worm crawling forward and a single-mode fit to the trail
(cf.\ Fig.\ \ref{Piece fit - trail}B).  We conclude that both
deviations result from the same general behavioral pattern, i.e., from
a sudden change in parameters of the harmonic-curvature wave
propagating through the body of \Celegans.

\section*{Discussion}

\input{figtex/swim_shapes}
\begin{figure}
\begin{picture}(270,160)
\putpic{0,160}{
\putpic{10,0}{
\includegraphics[width=0.32\textwidth, angle=270]{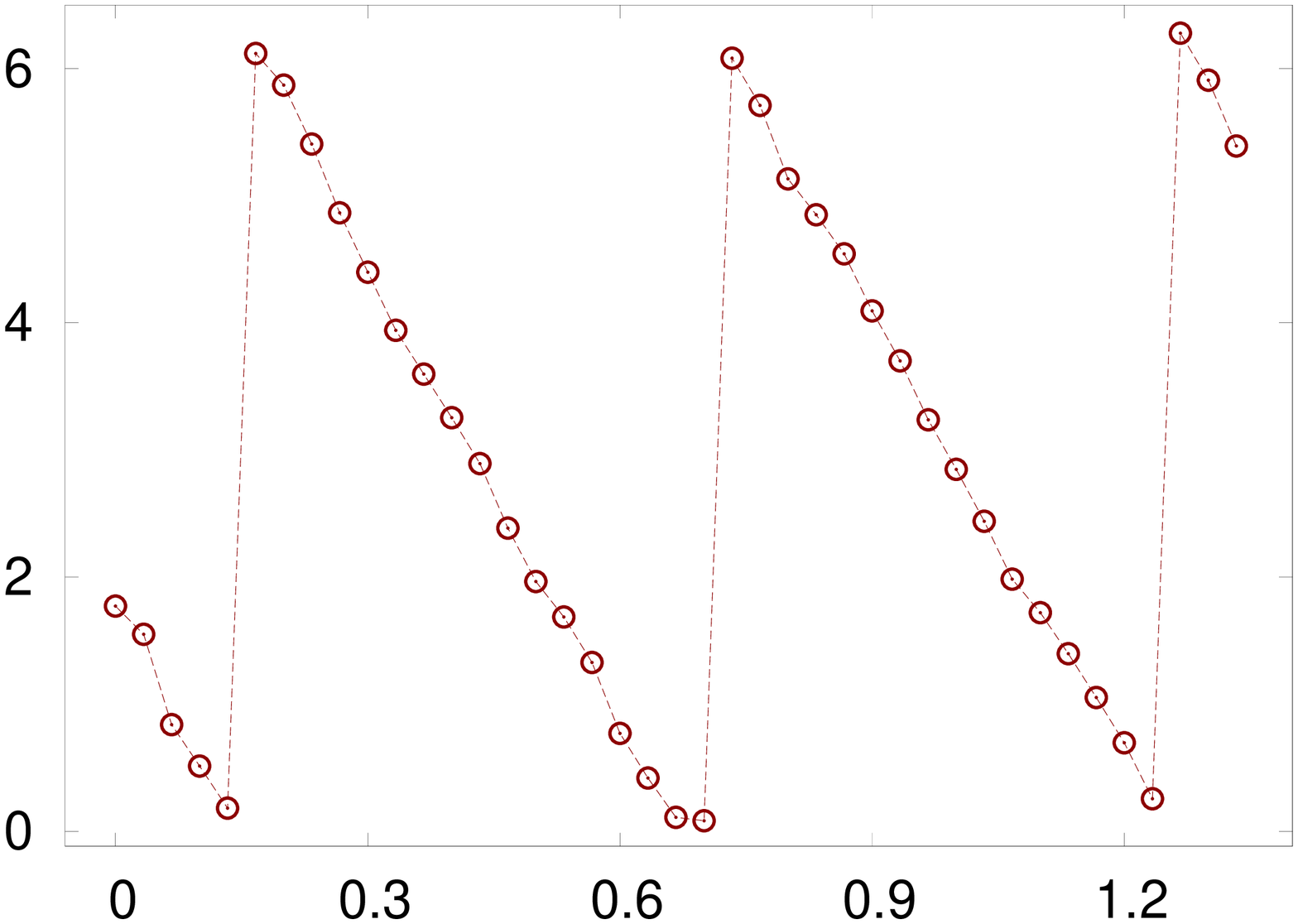}
}
\putpic{125,-161}{$t$ (sec)}
\putpic{14,-70}{$\phase$}
}
\end{picture}
\caption{{\bf The phase evolution $\phase$ for a sequence of images of
    \Celegans\ swimming in water.} The sequence of images from the
  movie of a swimming \Celegans\ was fitted using harmonic-curvature
  function \eqref{harmonic curvature} with the phase $\phi$ varying
  from frame to frame and the amplitude $\amplitude$ and wavevector
  $\wavevector$ identical for all frames.}
\label{Phase evolution}
\end{figure}

Our piecewise-harmonic-curvature model provides a simple analytical
description of the crawling motion of \Celegans\ in terms of
elementary sinusoidal movements in curvature space.  We have
demonstrated that the worm turns by changing the curvature-wave
parameters.  In this section we describe consequences of our findings
and indicate directions for further research.

\subsection*{Curvature evolution in the intrinsic worm coordinates}

Our observations provide important clues on how the neural system of
\Celegans\ controls the nematode movement to generate propulsion and
navigate the environment.  To elucidate this aspect, we transform our
curvature-based description to the intrinsic worm coordinates with the
arc length $\arcLengthWorm=\arcLength-\arcLengthTail(\timet)$ measured
from the position of the tip of the nematode tail, instead of the
beginning of the trail.  In this representation, the curvature
$\curvatureWorm$ at position $\arcLengthWorm$ and time $\timet$ can be
expressed in the form
\begin{equation}
\label{curvature wave}
\curvatureWorm(\arcLengthWorm,\timet)=\curvature(\arcLengthWorm
                                      +\wormVelocity\timet),
   \qquad 0\le\arcLengthWorm\le\wormLength,
\end{equation}
where $\curvature$ is the piecewise-harmonic function \eqref{harmonic
  curvature multiple}. To describe the tail position
$\arcLengthTail(\timet)$ we used Eq.\ \eqref{tail position time}, on
the assumption that the worm moves along the trail with a constant
velocity $\wormVelocity$.

\paragraphP{The worm's point of view on curvature modes}

According to our model, the worm motion is entirely determined by the
initial worm configuration at $\timet=0$,
\begin{equation}
\label{initial configuration}
\curvatureWorm(\arcLengthWorm,0)=\curvature(\arcLengthWorm),
   \qquad 0\le\arcLengthWorm\le\wormLength,
\end{equation}
and the boundary condition
\begin{equation}
\label{boundry condition}
\curvatureWorm(\wormLength,t)=\curvature(\wormLength+\wormVelocity\timet),
   \qquad \timet\ge0
\end{equation}
that describes the head curvature at time $t$.  The curvature
generated in the head segment propagates along the worm body according
to relation \eqref{curvature wave}.

It follows that the worm (i.e., the nervous system that controls its
body movements) needs to perform three, relatively simple,
independent tasks: 
\begin{enumerate}
\item[(i)] generate a harmonic oscillation of the curvature in the head
  part of the body, corresponding to the relation
\label{task 1}
\begin{equation}
\label{head oscillation}
\curvatureWorm(\wormLength,t)=\amplitude\sin(\frequency\timet+\phaseWorm),
\end{equation}
where $\frequency=\wavevector\wormVelocity$ is the angular frequency;
\item[(ii)] propagate the curvature wave backwards along the nematode, as
  described by Eq.\ \eqref{curvature wave};
\label{task 2}
\item[(iii)] produce random jumps of the amplitude $\amplitude$,
\label{task 3}
  wavevector $\wavevector$, and phase $\phaseWorm$ of the harmonic
  time variation in the head curvature \eqref{head
    oscillation}. 
\end{enumerate}
Since the curvature is an intrinsically local quantity, all three
tasks require only local operations.  Generation of the curvature in
the head segment (tasks (i) and (iii)) involves local muscle
contractions.  Propagation of the curvature wave (task (ii)) depends
on local muscle contractions and communication between neighboring
segments.

The simplicity of our geometrical description strongly suggests that
there is an underlying simplicity in the way \Celegans\ controls its
motion.  The worm is not aware of its entire posture, which would
require long-range communication between spatially separated body
segments, so the nematode controls its body locally.  The above
conjecture is also consistent with results of a recent study by
Stirman \textit{et al.\/}
\cite{Stirman-Crane-Husson-Wabnig-Schultheis-Gottschalk-Lu:2011}.  By
targeted illumination, these researchers have manipulated the behavior
of optogenetically engineered \Celegans.  In one of their experiments
(cf., Supplementary Video 1 of
\cite{Stirman-Crane-Husson-Wabnig-Schultheis-Gottschalk-Lu:2011})
neural excitation produced by a laser-light impulse applied to the
head section of a crawling nematode created a sharp bend in the worm
body.  After this initial perturbation, the bend propagated backwards
along the body of \Celegans, causing the nematode to follow the
direction of the head, and resulting in a worm crawling along an
artificially predetermined path (in the shape of a triangle).  This
behavior supports our conclusion that the propagation of the curvature
wave along the worm body is independent of the curvature-generation
mechanism.

\subsection*{Worms moving in different environments}

In this paper we have focused on $\Celegans$ moving without slip on a solid substrate.  However, based on our discussion of the way
the worm controls its body movements, we expect that key elements of
our description are likely to apply more generally, i.e., even when
\Celegans\ undergoes a significant slip, and the nematode trail is not
so well defined.

A typical example of such a situation is \Celegans\ swimming in water.
During swimming, the nematode undergoes a significant translational
and rotational drift with respect to the surrounding fluid.  Thus a
superposition of subsequent nematode images does not show a
well-defined trail.  To describe the motion of \Celegans\ in its
swimming gait, it is more convenient to analyze a sequence of
individual images (rather than their superposition) using intrinsic
worm coordinates, i.e., the framework defined by equation
\eqref{curvature wave}.

Two images from a video of a swimming \Celegans\ are shown in
Fig.\ \ref{Swim shapes} together with the harmonic-curvature fits to
the numerical skeletonization data. The nematode posture in each frame
can be accurately represented by contours corresponding to the
harmonic curvature \eqref{harmonic curvature} with a single mode of
the same amplitude and wavevector $\amplitude/\wavevector\approx0.96$
and $\wavevector\wormLength\approx4.83$, but a varying phase $\phi$.
This suggests that the internal kinematics of a swimming and crawling
\Celegans\ are quite similar. The evolution of phase $\phase$ as a
function of time for a sequence of video images is shown in
Fig. \ref{Phase evolution}. The gradual drop in $\phase$ with constant
$\amplitude$ and $\wavevector$ suggests that a single wave propagates
backward as the nematode swims.

A more detailed analysis is needed to determine whether \Celegans\ in
the swimming gait uses a similar strategy to turn as in the crawling
gait.  However, based on previous reports that the worm uses
$\Omega$-turns for changing swimming direction \cite{Hart:2006}, we
postulate that a swimming worm, similar to a crawling worm, abruptly
varies the harmonic-wave parameters at irregular intervals, which
results in a modification of the swimming course.

We believe that the framework based on harmonic-curvature-wave
assumptions \eqref{harmonic curvature multiple} and \eqref{curvature
  wave} will be useful in the analysis of worm propulsion and
maneuverability in water.  We also expect that a similar framework can
be applied in microstructured environments, such as soil or arrays of
microfluidic pillars.

\subsection*{Conclusions and future directions}

Our study has demonstrated a significant geometrical simplicity of the
crawling gait of \Celegans.  We have shown that \Celegans\ propels
itself using a simple set of elementary movements, and that both
shallow turns and sharp loop- and $\Omega$-turns correspond to the
same geometrical family of contours generated by a harmonic variation
of the curvature along the body of the nematode.  The geometrical
simplicity revealed by our analysis directly reflects key underlying
aspects of the neural control of the nematode motion (as explained
above).

Our analytic description of the nematode kinematics provides a
powerful tool for quantitative investigations of nematode behavior.
For example, behavioral changes of \Celegans\ in response to chemical
signals can be quantified by the statistics of jumps of the
curvature-wave parameters $\amplitude$, $\wavevector$, and $\phase$.
Since the worm controls these parameters directly (in contrast to the
angles of the corresponding turns), further studies based on our
description are likely to reveal important information on underlying
mechanisms of chemotaxis.

We expect that our analysis will facilitate development of advanced
models of neural control of nematode motion. Current models
\cite{Boyle-Bryden-Cohen:2008} capture gross qualitative features of
the crawling gait of \Celegans\ fairly well. However, as implied by
our results, more comprehensive models should include separate
neural-network functional units that would (i) control curvature-wave
generation in the head segment of the worm, (ii) propagate the wave
along the worm body, and (iii) change the wave
parameters. Introduction of such models would be an important step
towards back-engineering of the neural control system of
\Celegans. Such models would also aid development of a control system
for future synthetic worm-like particles that can crawl autonomously.

\section*{Materials and Methods}

\subsection*{Experimental Details}
\label{experimental details}

\paragraphP{Worm Preparation}
\label{worm preparation}
\Celegans\ were cultured on a 60$\times$15 mm petri dish containing 4
\textit{wt}\% agar at 20$^\circ$C. The worms were fed every 4 days
with 0.05 ml of E. coli, and the petri dish was subsequently wrapped
with paraffin film to prevent the growth of other bacteria or mold. To
conduct experiments, a chunk of agar containing worms was transferred
onto a fresh bacteria-free 4 \textit{wt}\% agar plate
\cite{Stiernagle:2006}. Images were acquired once all the worms had
left the chunk and had begun crawling on the fresh plate.

\paragraphP{Image Acquisition}
\label{image acquisition}
Images of \Celegans\ crawling on agar plates were taken using a Zeiss
Stemi 2000-C stereo microscope with a StreamView CCD camera
(640$\times$480 pixels with a 7.4 $\mu$m$^2$ pixel size and an 8-bit
mono sensor), typically acquiring between 10-30 frames per second, and
were saved as black and white avi movie files. The agar plates were
illuminated from below by mounting the plates on one face of a right
angle prism with an aluminized hypotenuse (50 mm leg length, 70.7 mm
hypotenuse length). A Chiu Technical Corporation Lumina FO-150 fiber
optic illuminator with a focusing lens and light diffuser were used
to illuminate one right-angle face of the prism; the light was
directed through the sample by the hypotenuse face of the prism, and
was acquired by the stereo microscope. Worms imaged with this
apparatus appear as a dark feature on a light background.

\paragraphP{Image Analysis}
\label{image analysis}
Individual worm shapes were obtained from the avi movie files by
converting the movie to 8-bit black and white tiff images using {\sc
  ImageJ} software (http://rsbweb.nih.gov/ij/). The individual images
were then analyzed as follows. A greyscale threshold was applied to
the images to separate the worm from the background, small non-worm
objects were removed from the images based on their size, 'holes' in
the thresholded worms' bodies were filled and the solid binary images
of the worms' bodies were transformed into a skeleton running down the
center of the body by applying a morphological thinning
operation. These operations were carried out using the {\sc Matlab}
Image Processing toolbox routines {\it bwareaopen.m}, {\it imfill.m},
and {\it bwmorph.m} respectively. Individual data points were
collected from the resulting binary skeleton image by determining
which pixels had a value equal to 1, collecting their row and column
positions, and utilizing these positions as Cartesian $(x,y)$
coordinates in order to perform curvature fits. In the case of
Fig. \ref{Loop turn}, the above described method of automatic
skeletonization using the image processing software did not provide
analyzable skeleton data due to the overlapping body segments of the
worm. Therefore, in this case, the image processing software was
used to obtain the skeleton of the worm except in the overlapping
region. For the overlapping region, the position of the pixels along
the centroid of the worm were selected and joined with the rest of the
worm body.

Worm trajectories, such as those shown in Figs. \ref{Piece fit -
  trail}, \ref{Piece fit 651}, and \ref{Piece fit 34}, were
constructed by opening an avi movie of crawling worms in {\sc ImageJ},
and projecting the darkest pixels from the time-series of images into
a single image (using {\sc ImageJ}'s {\it Z project} tool). This
resulting image was then processed in the same manner as individual
worm images in order to obtain $(x,y)$ data points along the worms'
trajectories.

\subsection*{Modeling details}
\label{modeling details}

\paragraphP{Solution of differential equations}
\label{solution of differential equations}
The set of coupled differential equations \eqref{explicit
  Frenet-Serret equations} was solved using the {\sc Matlab}'s
fourth-order Runge-Kutta ODE solver {\it ode45.m}. Since
Eq. \eqref{explicit Frenet-Serret equations} is a set of second order
differential equations, it requires two initial conditions---the
starting point and the initial direction. These two conditions were
obtained by the fitting procedure described below. For the continuous
representation with multiple modes along the trajectory, each mode was
determined using the location and direction of the last point of the
previous mode as the initial conditions. With sudden jumps in
parameters for the harmonic-curvature function, the curvature is
discontinuous between each mode as shown in Fig.\ \ref{Piece fit -
  parameters}B. However, the corresponding PHC representation of the
trajectory in real (Cartesian) space is a continuous curve with a
continuous slope, because it is a solution of a set of second-order
differential equations.

\paragraphP{Measure of error}
\label{measure of error}
The deviation of the model from the experimental data was measured
using the error function
\begin{equation}
\label{error function}
\epsilon_p=\wormLength^{-1}\left(\frac{1}{n}\sum_{i=1}^{n} d_i^p \right)^\frac{1}{p},
\end{equation}
where $n$ is the number of data points, and $d_i$ is the distance of
the curve \eqref{parametric body shape} from the data point {\it
  i}. The quantity \eqref{error function} was minimized to obtain the
model parameters for accurate description of the worm body postures
and trajectories. In all cases except one we used $p$ = 2. For the
trail presented in Fig. \ref{Piece fit 651}A, we used $p$ = 12 to
avoid large local deviations between the model and the data. The error
\eqref{error function} was minimized using the built-in {\sc Matlab}
optimizer {\it fminsearch.m}.

The local errors
\begin{equation}
\label{local error function}
\epsilon\left(\arcLength_j\right)=\wormLength^{-1}\left(\frac{1}{5}\sum_{i=j-2}^{j+2} d_i^2 \right)^\frac{1}{2},
\end{equation}
shown in the insets of Figs. \ref{Worm shapes 2} and \ref{Piece fit -
  trail} were estimated by calculating the average distance of the
curve from the data along a moving window of five data points.

\enlargethispage{36pt}

\paragraphP{Model parameters $\amplitude_i$, $\wavevector_i$, and
  $\phase_i$}
\label{fitting procedure}
The model parameters for a given trajectory were evaluated in two
stages. In the first stage, individual segments of the trajectory were
fitted independently using the single-mode representation
\eqref{harmonic curvature} to find the positions of the mode shifts
and to determine the initial guesses for the parameters
$\amplitude_i$, $\wavevector_i$, and $\phase_i$. In the second stage,
a continuous multi-mode representation of the trajectory was obtained
using the initial guesses from the first step.

{\it Single-mode independent piecewise fits}.  The fitting procedure
was started from one end of the trail with approximately 25
consecutive skeleton points. The model parameters for this section
were obtained by minimizing Eq. \eqref{error function}. The quality of
the fit was estimated by calculating the local deviation between the
model and the experimental data using Eq. \eqref{local error
  function}. If the local error of a particular segment showed no
increase, the length of the section was doubled and the above steps
were repeated. If the local error showed a steady increase beyond a
certain point, as seen in Fig. \ref{Piece fit - trail}B, the fit was
truncated at that point and the point of truncation served as the
location of a mode shift. The above steps of independent harmonic fits
were repeated until the end of the trajectory was reached.

{\it Continuous Fit}.  The purpose of this stage was to determine the
parameters $\amplitude_i$, $\wavevector_i$, and $\phase_i$ for the
continuous representation of a trajectory according to the PHC model
\eqref{harmonic curvature multiple}. The continuous representation was
obtained by consecutively fitting larger and larger portions of the
trajectory to avoid difficulties associated with local minima of the
accuracy measure \eqref{local error function}. The fitting procedure
was started from the first segment and involved addition of subsequent
segments in two steps. In the first step, the curvature parameters of
segment {\it k} were refitted with the constraints of maintaining the
continuity of the curve and its slope between the current and previous
segments. In the second step, segments 1 through {\it k} were
readjusted allowing variations of their curvature parameters and
locations of mode shifts. Typically, we allow only variations of
parameters of segments {\it k}-2, {\it k}-1, and {\it k} in this
step. The above two steps were repeated until the end of the
trajectory was reached.

In several cases (Figs. \ref{Piece fit - trail}, \ref{Piece fit 651}B,
and \ref{Piece fit 34}A), the above procedure did not provide an
accurate match between the model and the trajectory. This difficulty
was associated with defects of the trail due to the local slip of the
worm. For these cases, the continuous PHC representation of the
trajectories was obtained by first fitting the two sections (before
and after the region with imperfections) independently, and then
joining them smoothly using an additional segment corresponding to the
region of the imperfect data.

In all cases, we achieved the accuracy
$\epsilon\left(\arcLength_j\right)\lesssim$0.02 and the typical
segment length was 0.7$\wormLength$. The fits performed in the forward
and reverse directions yielded similar parameter values, which
indicates that our model is robust.

\section*{Acknowledgments}
We would like to thank Frank Van Bussel for valuable discussions and
programming advice and Alejandro Bilbao for his contribution to
initial observations of swimming \Celegans.  SAV acknowledges NSF
CAREER Award Grant No.\ 1150836. This work was also supported by NSF
Grant No.\ CBET 1059745 (JB,VP).

\section*{Author Contributions}
Conceived and designed the project: JB SAV KPR VP.
Designed the experimental setup: ZSK.
Performed the experiments: VP DES ZSK.
Analyzed the data: ZSK VP.
Contributed materials: AA.
Wrote the paper: VP JB.

\bibliographystyle{unsrt} 
\bibliography{vbib}

\end{document}